\documentclass[12pt,preprint,onecolumn]{aastex}

\usepackage{graphicx}
\usepackage{epsfig}
\usepackage{natbib}
\usepackage{hyperref}
\usepackage{bm}
\usepackage{amsmath}

\renewcommand{\vec}[1]{\bm{#1}}

\begin{document}

\title{CLASP Constraints on the Magnetization and Geometrical Complexity of \\the Chromosphere-Corona Transition Region}

\author{
J.~Trujillo Bueno\altaffilmark{1,2,3},
J.~\v{S}t\v{e}p\'an\altaffilmark{4,1},
L.~Belluzzi\altaffilmark{5,6},
A.~Asensio Ramos\altaffilmark{1},
R.~Manso Sainz\altaffilmark{7},
T.~del Pino Alem\'an\altaffilmark{1},
R.~Casini\altaffilmark{8},
R.~Ishikawa\altaffilmark{9},
R.~Kano\altaffilmark{9},
A.~Winebarger\altaffilmark{10},
F.~Auch\`ere\altaffilmark{11},
N.~Narukage\altaffilmark{9},
K.~Kobayashi\altaffilmark{10},
T.~Bando\altaffilmark{9},
Y.~Katsukawa\altaffilmark{9},
M.~Kubo\altaffilmark{9},
S.~Ishikawa\altaffilmark{12},
G.~Giono\altaffilmark{9},
H.~Hara\altaffilmark{9},
Y.~Suematsu\altaffilmark{9},
T.~Shimizu\altaffilmark{12},
T.~Sakao\altaffilmark{12},
S.~Tsuneta\altaffilmark{9},
K.~Ichimoto\altaffilmark{9,13},
J.~Cirtain\altaffilmark{10},
P.~Champey\altaffilmark{10},
B.~De~Pontieu\altaffilmark{14,15,16},
and
M.~Carlsson\altaffilmark{15,16}
}
\altaffiltext{1}{Instituto de Astrof\'isica de Canarias, E-38205 La Laguna, Tenerife, Spain}
\altaffiltext{2}{Universidad de La Laguna, Departamento de Astrof\'isica, 38206 La Laguna, Tenerife, Spain}
\altaffiltext{3}{Consejo Superior de Investigaciones Cient\'ificas, Spain}
\altaffiltext{4}{Astronomical Institute ASCR, Fri\v{c}ova 298, 251\,65 Ond\v{r}ejov, Czech Republic}
\altaffiltext{5}{Istituto Ricerche Solari Locarno, CH - 6605 Locarno Monti, Switzerland}
\altaffiltext{6}{Kiepenheuer-Institut f\"ur Sonnenphysik, D-79104 Freiburg, Germany}
\altaffiltext{7}{Max-Planck-Institut f\"ur Sonnensystemforschung, Justus-von-Liebig-Weg 3, 37077 G\"ottingen, Germany}
\altaffiltext{8}{High Altitude Observatory, National Center for Atmospheric Research, Post Office Box 3000, Boulder, CO 80307-3000, USA}
\altaffiltext{9}{National Astronomical Observatory of Japan, National Institutes of Natural Science, 2-21-1 Osawa, Mitaka, Tokyo 181-8588, Japan}
\altaffiltext{10}{NASA Marshall Space Flight Center, ZP 13, Huntsville, AL 35812, USA}
\altaffiltext{11}{Institut d'Astrophysique Spatiale, B\^atiment 121, Univ. Paris-Sud - CNRS, 91405 Orsay Cedex, France}
\altaffiltext{12}{Institute of Space and Astronautical Science, Japan Aerospace Exploration Agency, 3-1-1 Yoshinodai, Chuo, Sagamihara, Kanagawa 252-5210, Japan}
\altaffiltext{13}{Hida Observatory, Kyoto University, Takayama, Gifu 506-1314, Japan}
\altaffiltext{14}{Lockheed Martin Solar and Astrophysics Laboratory, Palo Alto, California, USA}
\altaffiltext{15}{Institute of Theoretical Astrophysics, University of Oslo, P.O. Box 1029 Blindern, NO-0315 Oslo, Norway}
\altaffiltext{16}{Rosseland Centre for Solar Physics, University of Oslo, P.O. Box 1029 Blindern, NO-0315 Oslo, Norway}

\shorttitle{Probing the Chromosphere-Corona Transition Region}

\shortauthors{Trujillo Bueno et al.}

\begin{abstract}

The Chromospheric Lyman-Alpha Spectro-Polarimeter (CLASP) is a suborbital rocket experiment that on 3rd September 2015 measured the linear polarization produced by scattering processes in the hydrogen Ly-$\alpha$ line of the solar disk radiation, whose line-center photons stem from the chromosphere-corona transition region (TR). These unprecedented spectropolarimetric observations revealed an interesting surprise, namely that there is practically no center-to-limb variation (CLV) in the $Q/I$ line-center signals. Using an analytical model, we first show that the geometrical complexity of the corrugated surface that delineates the TR has a crucial impact on the CLV of the $Q/I$ and $U/I$ line-center signals. Secondly, we introduce a statistical description of the solar atmosphere based on a three-dimensional (3D) model derived from a state-of-the-art radiation magneto-hydrodynamic simulation. Each realization of the statistical ensemble is a 3D model characterized by a given degree of magnetization and corrugation of the TR, and for each such realization we solve the full 3D radiative transfer problem taking into account the impact of the CLASP instrument degradation on the calculated polarization signals. Finally, we apply the statistical inference method presented in a previous paper to show that the TR of the 3D model that produces the best agreement with the CLASP observations has a relatively weak magnetic field and a relatively high degree of corrugation. We emphasize that a suitable way to validate or refute numerical models of the upper solar chromosphere is by confronting calculations and observations of the scattering polarization in ultraviolet lines sensitive to the Hanle effect. 
\end{abstract}

%\date{Received XXXX; accepted XXXX}

\keywords{Sun: transition region --- Sun: UV radiation --- Magnetic fields --- Polarization }

\newpage

%%%%%%%%%%%%%%%%%%%%%%%%%%%%%%%%%%%%%%%%
%%%%%%%%%%%%%%%%%%%%%%%%%%%%%%%%%%%%%%%%
%%%%    SECTION 1
%%%%%%%%%%%%%%%%%%%%%%%%%%%%%%%%%%%%%%%%
%%%%%%%%%%%%%%%%%%%%%%%%%%%%%%%%%%%%%%%%

\section{Introduction}
\label{sec:S1}

Recent theoretical investigations 
predicted that the hydrogen Ly-$\alpha$ line of the solar disk radiation
should be linearly polarized by the scattering of anisotropic radiation, with measurable polarization signals in both the core and wings of the line (Trujillo Bueno et al. 2011; Belluzzi et al. 2012; \v{S}t\v{e}p\'an et al. 2015). Moreover, such radiative transfer investigations pointed out that the line-center polarization is modified by the presence of magnetic fields in the chromosphere-corona TR via the Hanle effect.
These theoretical investigations motivated the development of the Chromospheric Lyman-Alpha SpectroPolarimeter (CLASP), an international sounding rocket experiment that on 3rd September 2015 successfully measured the wavelength variation of the intensity and linear polarization of the Ly-$\alpha$ line in quiet regions of the solar disk (see Kano et al. 2017, and references therein to the papers describing the instrument). 

\begin{figure}[t]
\begin{center}
\includegraphics[scale=0.9]{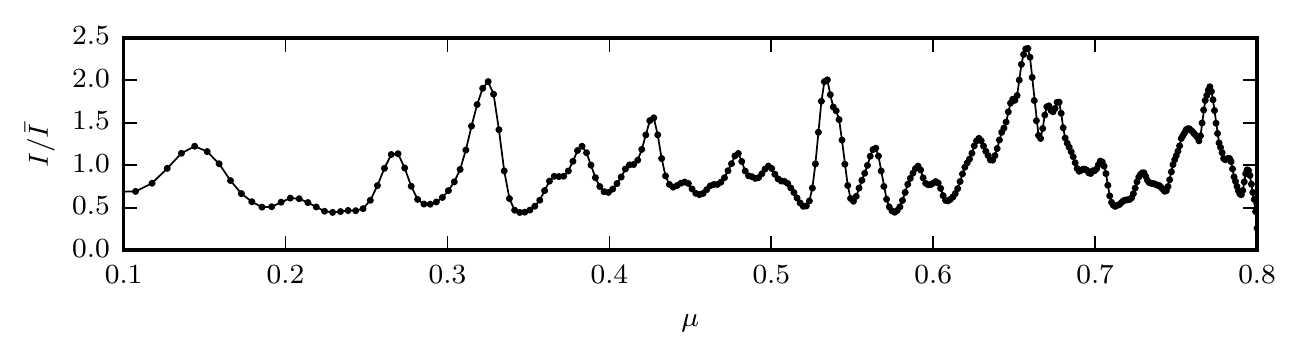}
\includegraphics[scale=0.9]{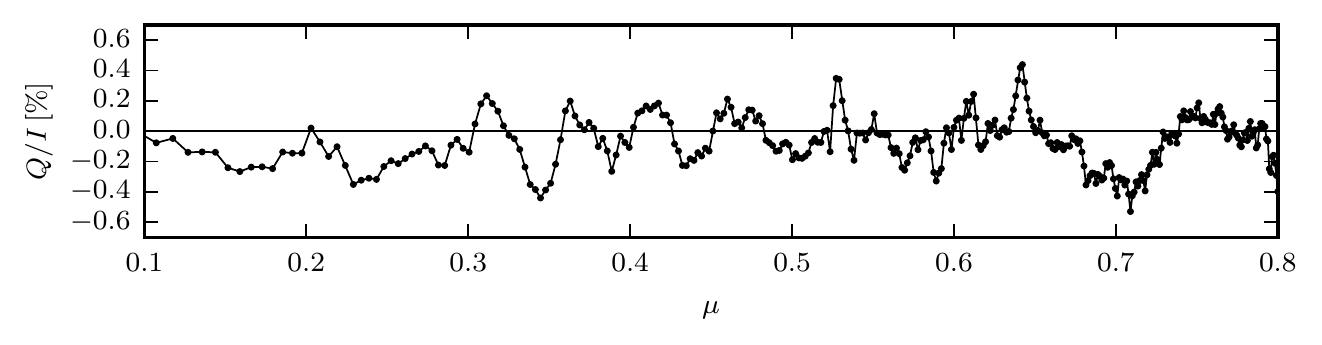}
\includegraphics[scale=0.9]{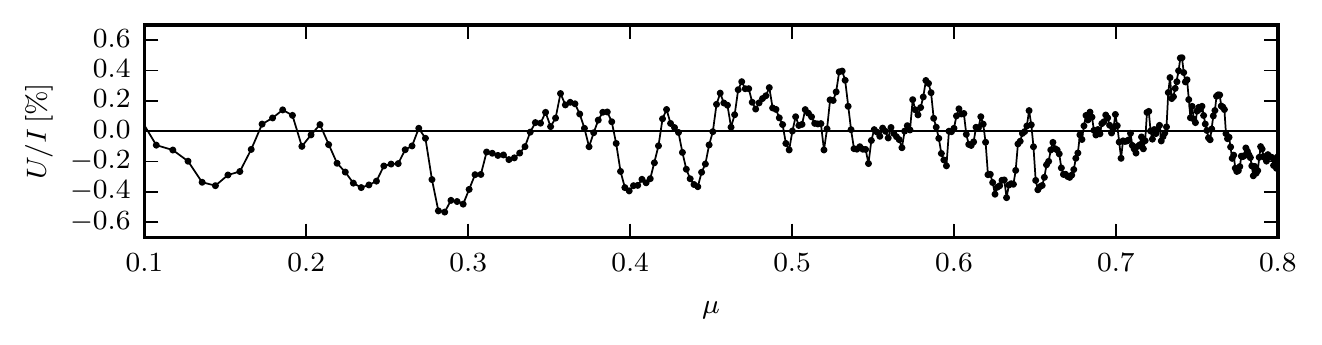}
\caption{CLASP observations.  
The variation with the cosine of the heliocentric angle ($\mu$) 
of the Stokes $I$, $Q/I$ and $U/I$ line-center signals observed by CLASP in the hydrogen Ly-$\alpha$ line. 
The reference direction for the positive Stokes $Q$ is the parallel to the nearest limb.}
\label{figure-1}
\end{center}
\end{figure}

CLASP observed the Ly-$\alpha$ Stokes profiles $I$, $Q$, and $U$ with a spatial resolution of about 3 arcsec and a temporal resolution of 5 minutes. The resulting $Q/I$ and $U/I$ linear polarization signals are of order 0.1\% in the line center and up to a few percent in the nearby wings, both showing conspicuous spatial variations with scales of ${\sim}10$ arcsec, in agreement with the theoretical predictions. However, the observed $Q/I$ line-center signals do not show any clear center-to-limb variation (CLV; see Fig. \ref{figure-1}), in contrast with the results of the above-mentioned radiative transfer investigations in several models of the solar atmosphere. 
The observed lack of a clear CLV in the $Q/I$ line-center signal came as an interesting surprise because many other solar spectral lines show such a variation (e.g., Stenflo et al. 1997), including the K line of Ca {\sc ii} whose line-center $Q/I$ and $U/I$ spatial variations are sensitive to the magnetic and thermal structure of the chromosphere at heights a few hundred kilometers below the TR (Holzreuter \& Stenflo 2007).   

The plasma of the upper solar chromosphere is highly structured and dynamic, departing radically from a horizontally homogeneous model atmosphere. In 3D models resulting from magneto-hydrodynamic (MHD) simulations (e.g., Carlsson et al. 2016), the TR,
where the temperature suddenly increases from about $10^4$ K to the $10^6$ K coronal values, is not a thin horizontal layer, but it delineates a highly corrugated surface.
In this work, such corrugated surface is characterized by specifying, at each point, the vector $\vec n$ indicating the direction of the local temperature gradient (hereafter, the local TR normal vector). 

We begin by illustrating that the spatial variations of the line-center $Q/I$ and $U/I$ signals of the Ly-$\alpha$ line are very sensitive to the geometric complexity of the corrugated surface that delineates the TR. Secondly, we demonstrate that the significant CLV of the $Q/I$ line-center signal of the Ly-$\alpha$ line 
calculated in Carlsson et al's (2016) 3D radiation MHD model of the solar atmosphere can be reduced by increasing the magnetic field strength and/or the geometrical complexity of the model's TR. We then show how this can be exploited to constrain, from the CLASP line-center data, the magnetic strength and geometric complexity of the solar TR by applying the statistical inference method discussed in \v{S}t\v{e}p\'an et al. (2018). To this end, we confront the statistical properties of the CLASP line-center data with those of the polarization signals calculated in a grid of 3D model atmospheres characterized by different degrees of geometrical complexity and magnetization. 

%%%%%%%%%%%%%%%%%%%%%%%%%%%%%%%%%%%%%%%%
%%%%%%%%%%%%%%%%%%%%%%%%%%%%%%%%%%%%%%%%
%%%%    SECTION 2
%%%%%%%%%%%%%%%%%%%%%%%%%%%%%%%%%%%%%%%%
%%%%%%%%%%%%%%%%%%%%%%%%%%%%%%%%%%%%%%%%

\section{An analytical corrugated transition region model}
\label{sec:S2}

In a 1D model atmosphere, static or with radial velocities, unmagnetized or with
a magnetic field having a random azimuth at sub-resolution scales, 
the radiation field at each point within the medium has axial symmetry around the vertical.
Under such circumstances, taking a reference system with the Z-axis (i.e., the quantization axis) along the vertical, and choosing the reference direction for linear polarization perpendicular to the plane formed by the vertical and the LOS (i.e., the parallel to the limb), then $U/I=0$ and the following approximate formula can be applied to estimate the line-center $Q/I$ scattering polarization signal of the hydrogen Ly-$\alpha$ line (Trujillo Bueno et al. 2011):

\begin{equation}
\frac QI\,\approx\,{1\over{2\sqrt{2}}}(1-\mu^2)\,{\cal H}\,{{{\bar J}^2_0}\over{{\bar J}^0_0}}\,,
\label{eq:qieddb}
\end{equation}
where $\mu={\rm cos}\,{\theta}$ (with $\theta$ the heliocentric angle), ${\cal H}$ is the Hanle depolarization factor (${\cal H}=1$ for $B=0$ G and ${\cal H}{<}1$ 
in the presence of a  magnetic field), and ${{\bar J}^2_0}/{{\bar J}^0_0}$ is the degree of anisotropy of the spectral line radiation at the height in the model atmosphere where the line-center optical depth is unity along the line of sight (see equations 2 and 3 of Trujillo Bueno et al. 2011). As seen in figure 1 of Trujillo Bueno et al. (2011), the anisotropy factor of the hydrogen Ly-$\alpha$ line suddenly becomes significant at the atmospheric height where the line-center optical depth is unity, which practically coincides with the location of the model's chromosphere-corona TR. Therefore, under such assumptions, we should expect a clear CLV in the $Q/I$ line-center signal, established by the $(1-\mu^2)$ factor of Eq. (1), and modified by the height-dependence of the anisotropy factor (see the height range between the two solid-line arrows in figure 1 of Trujillo Bueno et al. 2011). Because of our choice for the reference direction for linear polarization, the only way to have $U/I{\ne}0$ in a static 1D model atmosphere is by means of the Hanle effect of a magnetic field inclined with respect to the local vertical. However, in such a 1D model atmosphere there is no way to destroy the CLV of the $Q/I$ line-center signal.

The line-center photons of the hydrogen Ly-$\alpha$ line originate just at the boundary of the model's TR.
In a 1D model atmosphere (e.g., Fontenla et al. 1993), the vector normal to the TR lies along the vertical, which coincides with the symmetry axis of the incident radiation field. In a three-dimensional model atmosphere (e.g., Carlsson et al. 2016), the model's TR delineates a corrugated surface, so that the vector normal to the model's TR changes its direction from point to point (see figure 7 of \v{S}t\v{e}p\'an et al. 2015). On the other hand, at each point on such corrugated surface the stratification of the physical quantities along the local normal vector $\vec n$ is much more important than along the perpendicular direction. In order to estimate how the line-center fractional polarization of the hydrogen Ly-$\alpha$ line is at each point of the field of view, it is reasonable to assume that the incident radiation field has axial symmetry around the direction of the normal vector $\vec n$ corresponding to the spatial point under consideration (see also \v{S}t\v{e}p\'an et al. 2018). Taking a reference system with the Z-axis directed along the normal vector $\vec n$, and recalling Eq. (1), at each point of the corrugated TR surface we can estimate the line-center fractional polarization signals through the following formula
  
\begin{equation}
{\Big{[}}\frac QI{\Big{]}}_{\vec n}\,\approx\,{1\over{2\sqrt{2}}}(1-{\mu}_{\vec n}^2)\,{\cal H}\,{\Big{[}}{{{\bar J}^2_0}\over{{\bar J}^0_0}}{\Big{]}}_{\vec n} \, ,
\label{eq:qieddb}
\end{equation}
where the positive Stokes $Q$ reference direction is now the perpendicular to the plane formed by $\vec n$ and the LOS, ${\mu}_{\vec n}$ is the cosine of the angle between $\vec n$ and the LOS, and ${{[}}{{\bar J}^2_0}/{\bar J}^0_0{{]}}_{\vec n}$ is the anisotropy factor calculated in the new reference system. 
Clearly, since the direction of $\vec n$ changes as we move through the corrugated surface that delineates the TR, the $Q/I$ signals estimated with Eq. (2) do not share the same reference direction for the quantification of the linear polarization. In order to arrive at equations for $Q/I$ and $U/I$ having the parallel to the nearest limb as the positive Stokes $Q$ reference direction, we have applied suitable rotations of the reference system. Specifying the direction of the local TR normal vector through its inclination $\theta_{\vec n}$ with respect to the vertical (with $\theta_{\vec n}$ between $0^{\circ}$ and $90^{\circ}$) and azimuth $\chi_{\vec n}$ (with $\chi_{\vec n}$ between $0^{\circ}$ and $360^{\circ}$), our analytical calculations show that the line-center $Q/I$ and $U/I$ signals of the hydrogen Ly-$\alpha$ line can be estimated by the following formulas: 

\begin{equation}
\frac QI\,\approx\,{\cal H}{1\over{\sqrt{2}}}\,{[}{1\over{4}}{\sin}^2{\theta}(3{\cos}^2{\theta_{\vec n}}-1)-{\sin}{\theta}{\cos}{\theta}{\sin}{\theta_{\vec n}}{\cos}{\theta_{\vec n}}{\cos}{\chi_{\vec n}}+{1\over{4}}(1+{\cos}^2{\theta})\,{\sin}^2{\theta_{\vec n}}{\cos}{(2\chi_{\vec n})}\,{]}\,[{{{\bar J}^2_0}\over{{\bar J}^0_0}}]_{\vec n},
\end{equation}

\begin{equation}
\frac UI\,\approx\,{\cal H}{1\over{\sqrt{2}}}\,[{1\over{2}}{\cos}{\theta}{\sin}^2{\theta_{\vec n}}{\sin}{(2\chi_{\vec n})}-{\sin}{\theta}{\sin}{\theta_{\vec n}}{\cos}{\theta_{\vec n}}{\sin}{\chi_{\vec n}}]\,[{{{\bar J}^2_0}\over{{\bar J}^0_0}}]_{\vec n}.
\end{equation}
Figure 2 shows examples of CLV of the $Q/I$ and $U/I$ Ly-$\alpha$ line-center signals for several topologies of the model's TR, with random azimuth $\chi_{\vec n}$. The first three rows of the figure correspond to the indicated fixed inclinations ${\theta_{\vec n}}$, while the bottom row panels 
show the case in which also ${\theta_{\vec n}}$ has random values (i.e., the probability distribution $p(\cos{\theta}_{\vec n})=1$ for $0{\le}\cos{\theta}_{\vec n}{\le}1$ and $p(\cos{\theta}_{\vec n})=0$ for $-1{\le}\cos{\theta}_{\vec n}{<}0$). 
Note that the CLV of the fractional linear polarization line-center signals is very sensitive to the geometry of the model's TR, and that there is no CLV at all when the normal vector $\vec n$ has random orientations. 
Moreover, it is also very important to point out that the $Q/I$ and $U/I$ amplitudes are sensitive to the magnetic field strength of the chromosphere-corona TR, through the Hanle factor $\cal H$ in Eqs. (3) and (4).

\begin{figure}[t]
\begin{center}
\includegraphics[scale=0.7]{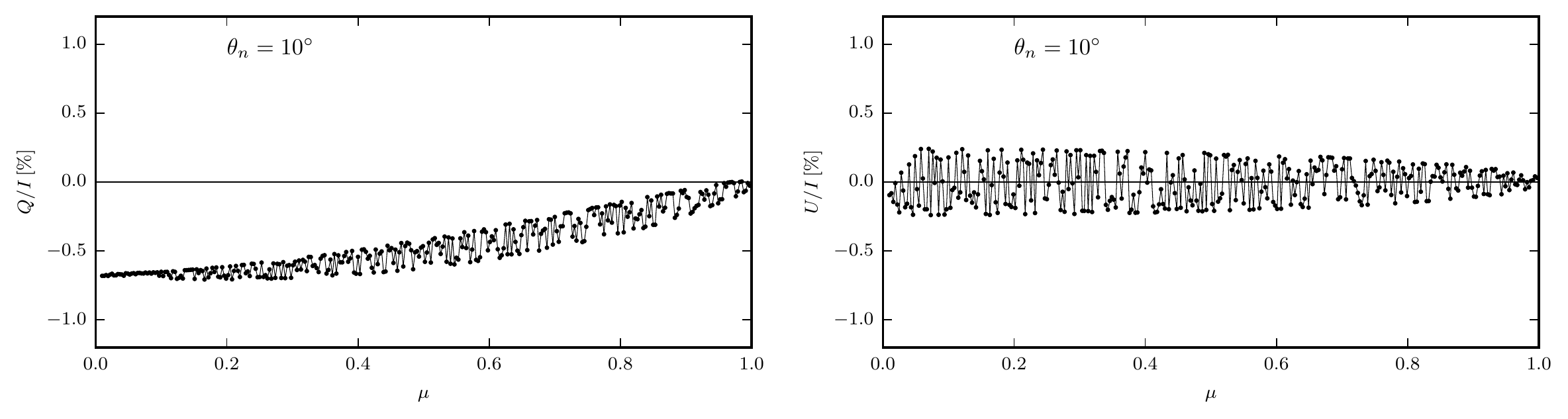}
\includegraphics[scale=0.7]{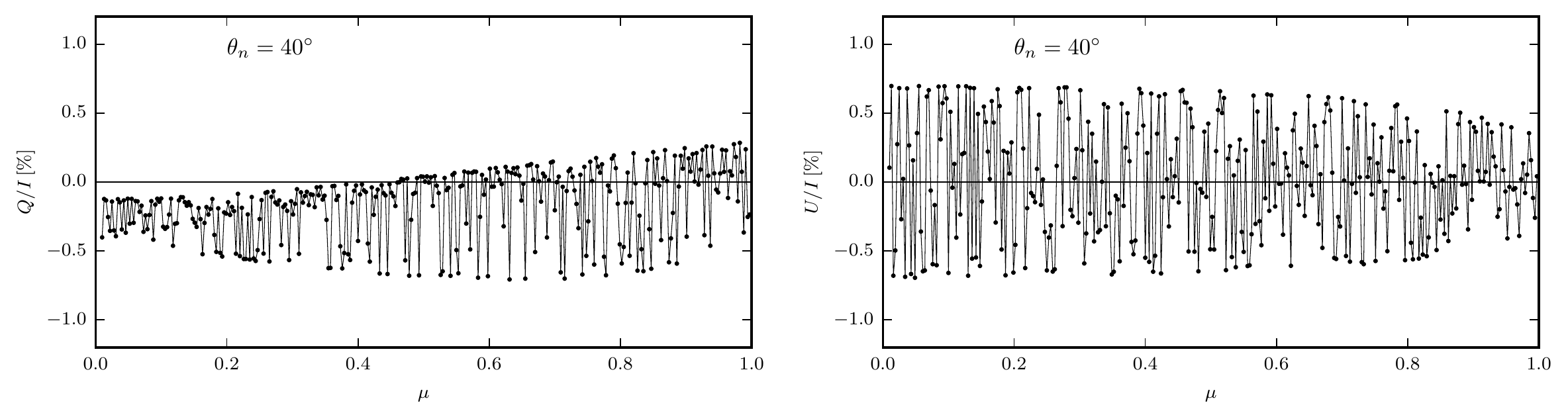}
\includegraphics[scale=0.7]{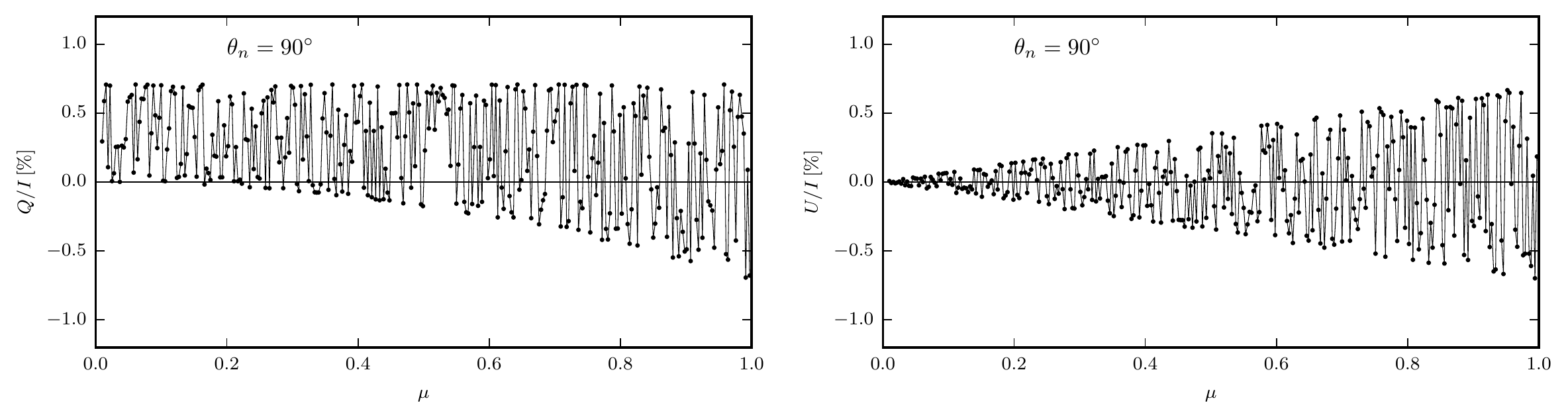}
\includegraphics[scale=0.7]{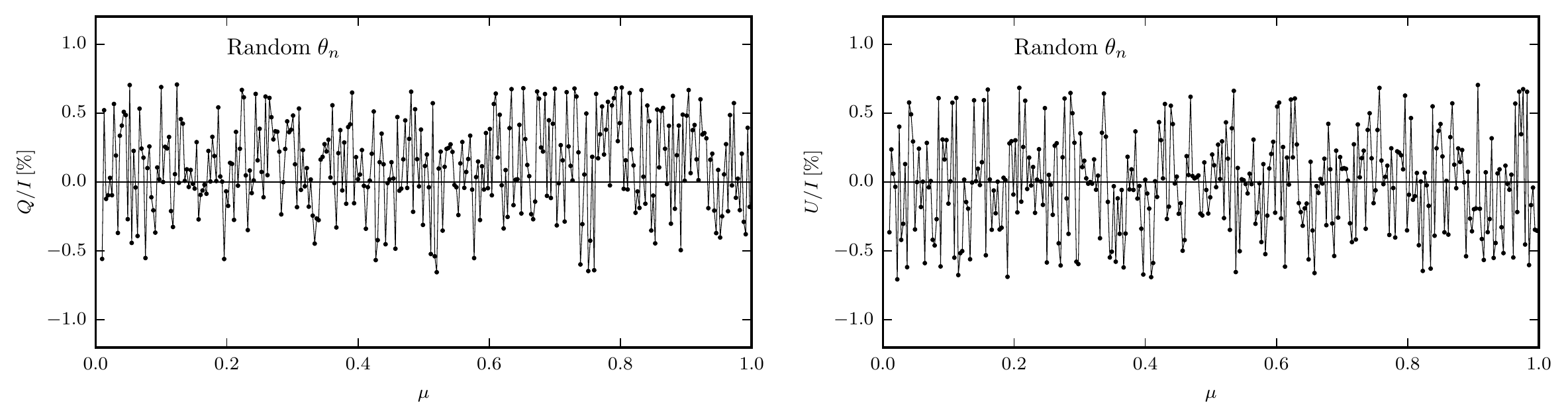}
\caption{The CLV of the Stokes $Q/I$ (left) and $U/I$ (right) line-center signals of the hydrogen Ly-$\alpha$ line calculated 
according to Eqs. (3) and (4) with ${\cal H}=1$ (no magnetic field) and ${\Big{[}{{{\bar J}^2_0}/{{\bar J}^0_0}}\Big{]}}_{\vec n}=-0.02$, 
for the fixed inclinations (${\theta_{\vec n}}$) of the TR normal vector $\vec n$ indicated in the panel insets and assuming random azimuths $\chi_{\vec n}$.  
Note that the bottom panel corresponds to the case in which ${\theta_{\vec n}}$ also has random values (see the text for details). 
The reference direction for the positive Stokes $Q$ is the parallel to the nearest limb.
}
\label{figure-2}
\end{center}
\end{figure}

%%%%%%%%%%%%%%%%%%%%%%%%%%%%%%%%%%%%%%%%
%%%%%%%%%%%%%%%%%%%%%%%%%%%%%%%%%%%%%%%%
%%%%    SECTION 3
%%%%%%%%%%%%%%%%%%%%%%%%%%%%%%%%%%%%%%%%
%%%%%%%%%%%%%%%%%%%%%%%%%%%%%%%%%%%%%%%%
\section{The impact of the magnetization and geometrical complexity in 3D models}
\label{sec:S3}

The analytical TR model described in the previous section is very useful to understand 
why the spatial variations of the $Q/I$ and $U/I$ line-center signals of the Ly-$\alpha$ line are very sensitive to the 
geometry of the corrugated surface that delineates the solar TR. The magnetic 
field of such analytical model is characterized by a random azimuth at sub-resolution scales, which implies that   
the impact of the model's magnetic field strength on the CLV is just the scaling factor $\cal H$ that appears in Eqs. (3) and (4). We now show theoretical results for the 3D radiation magneto-hydrodynamical model of the chromosphere-corona TR of Carlsson et al. (2016), which is representative of an enhanced network region and has magnetic field lines  reaching chromospheric and coronal heights. 

The first two rows of Fig. 3 show examples of the CLV of the $Q/I$ and $U/I$ line-center signals calculated 
with the radiative transfer code PORTA (\v{S}t\v{e}p\'an \& Trujillo Bueno 2013) in the 3D model of Carlsson et al. (2016) 
ignoring (first row) and taking into account (second row) the CLASP instrument degradation (see Giono et al. 2016 for information on the point spread function). 
Clearly, the peak-to-peak amplitudes of the $Q/I$ and $U/I$ variations are significantly smaller when the degradation produced by the instrument is accounted for. Note that the $Q/I$ line-center signals calculated in such a 3D model show a clear CLV. 

\begin{figure}[t]
\begin{center}
\includegraphics[scale=0.8]{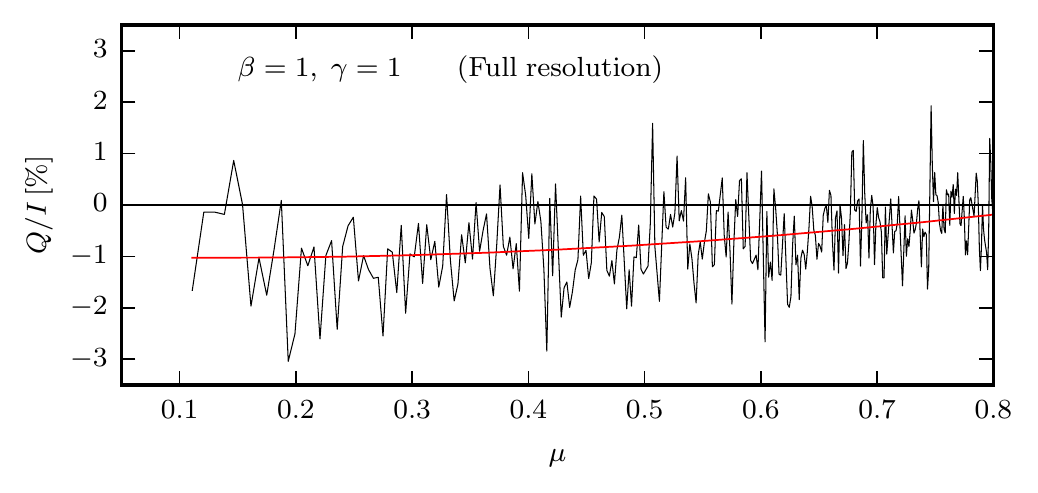}
\includegraphics[scale=0.8]{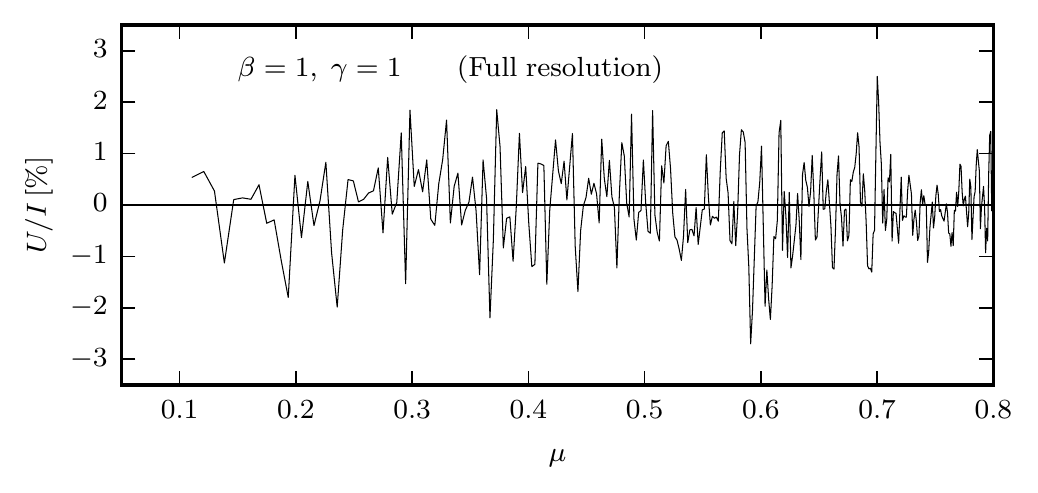}
\includegraphics[scale=0.8]{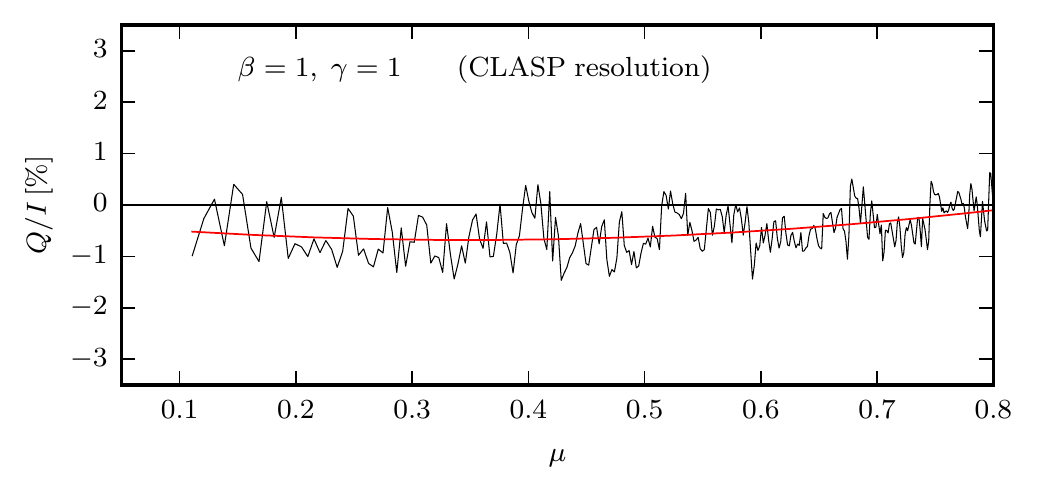}
\includegraphics[scale=0.8]{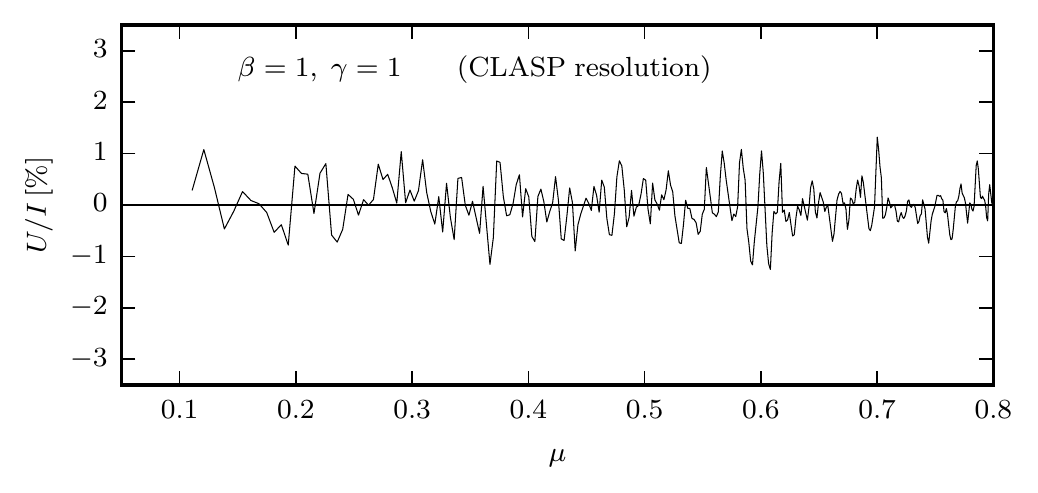}
\includegraphics[scale=0.8]{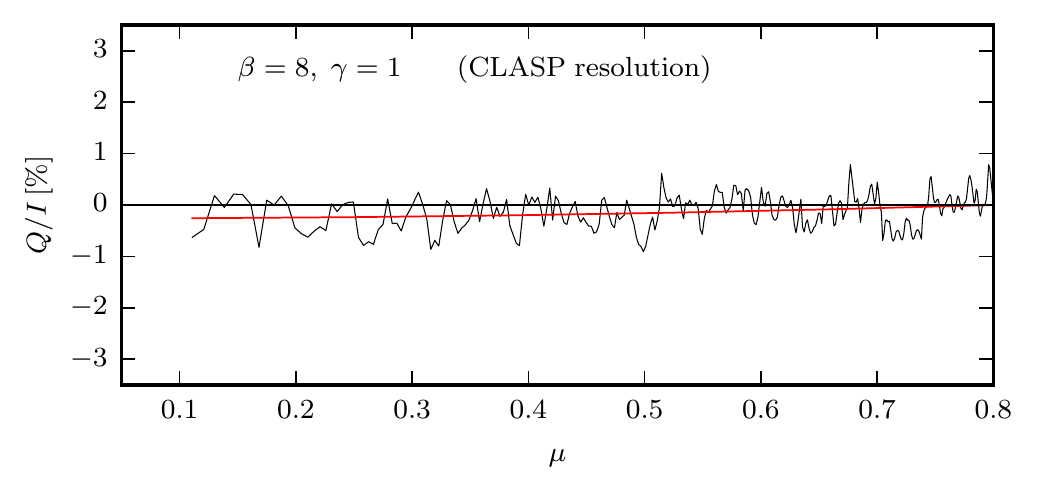}
\includegraphics[scale=0.8]{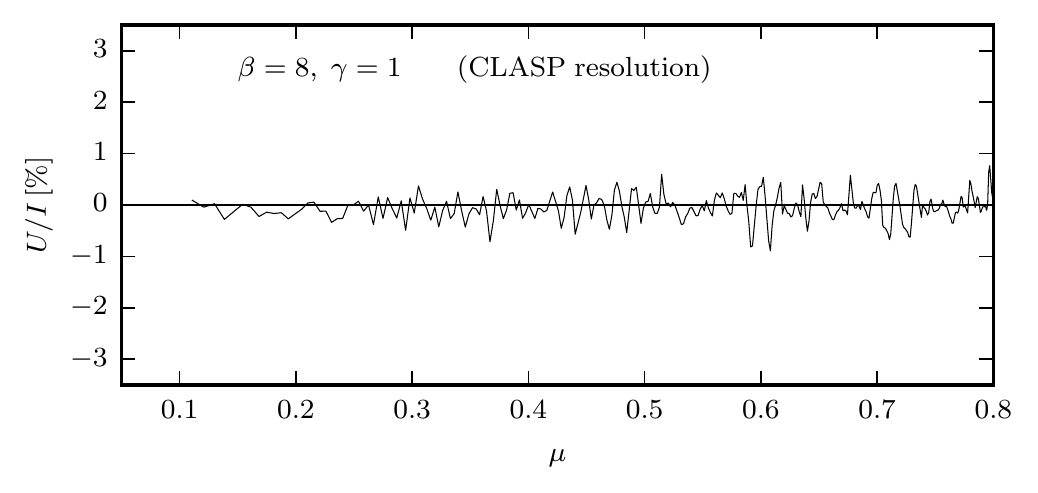}
\includegraphics[scale=0.8]{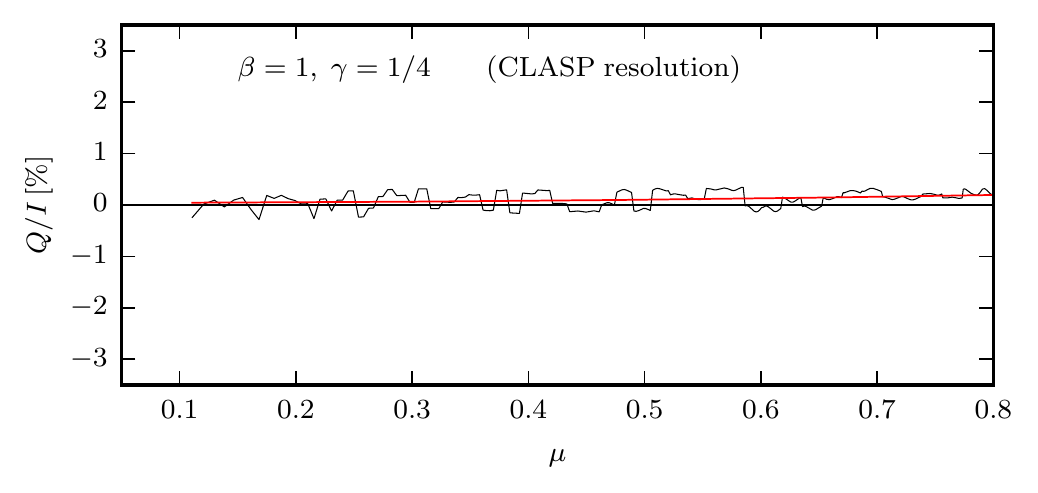}
\includegraphics[scale=0.8]{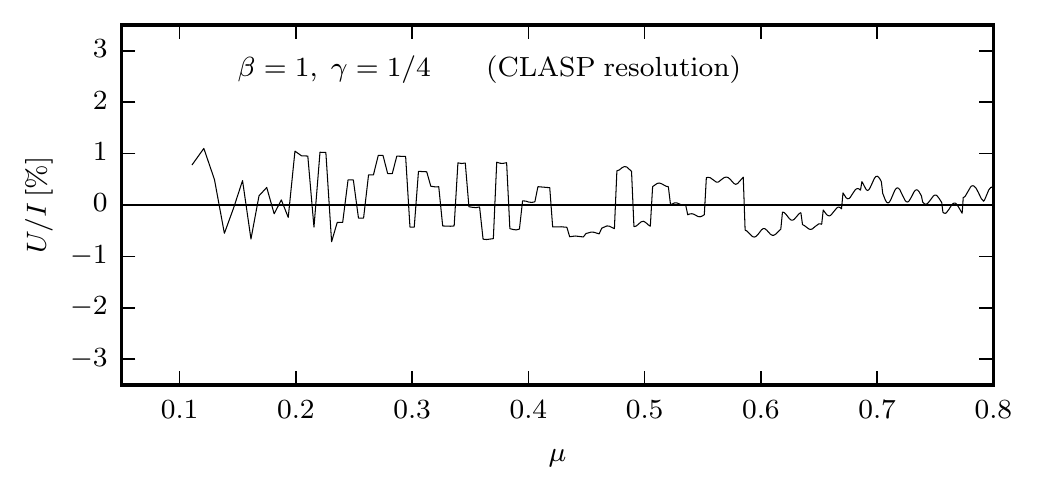}
\caption{The CLV of the Stokes $Q/I$ (left) and $U/I$ (right) line-center signals of the hydrogen Ly-$\alpha$ line calculated in the 3D model of Carlsson et al. (2016), ignoring (topmost panels) and accounting for (rest of the panels) the degradation produced by the CLASP instrument. 
The panels of the first and second rows show the results corresponding to the magnetic field and TR corrugation of the original 3D model. Those of the third row  
show the results obtained assuming an enhanced magnetization characterized by a scaling factor $\beta=8$ (which implies a mean field strength of 120 G in the model's TR). The panels of the fourth row show the results obtained assuming a compression factor $\gamma=1/4$ (which quantifies the geometrical complexity of the model's TR). The red curves in the left panels are best fits quadratic in $\mu$ to the fluctuating spatial variations. 
The reference direction for the positive Stokes $Q$ is the parallel to the nearest limb.
}
\label{figure-2}
\end{center}
\end{figure}

The panels in the third row of Fig. 3 show what happens when the magnetic field strength at each point in the 3D model is increased by a scaling factor $\beta=8$, so that the mean field strength at the model's TR is 120 G instead of 15 G. Clearly, an enhanced magnetization in the TR plasma has a significant impact on the linear polarization amplitudes, but it does not destroy the CLV of the $Q/I$ line-center signal. 

We show now what happens when the geometric complexity of the TR of the 3D model is modified. To this end, we simply compress the horizontal extension of the 3D model by a factor $\gamma$ along the $X$ and $Y$ directions, so that the divergence of the model's magnetic
field vector at each grid point remains equal to zero and its modulus unaltered. 
The panels in the bottom row of Fig. 3 shows an example for $\beta=1$ (i.e., the model's magnetic field), but with 
$\gamma=1/4$. Clearly, the degree of corrugation of the model's TR surface has an important impact on the CLV of the line-center $Q/I$ signals. 

%%%%%%%%%%%%%%%%%%%%%%%%%%%%%%%%%%%%%%%%
%%%%%%%%%%%%%%%%%%%%%%%%%%%%%%%%%%%%%%%%
%%%%    SECTION 4
%%%%%%%%%%%%%%%%%%%%%%%%%%%%%%%%%%%%%%%%
%%%%%%%%%%%%%%%%%%%%%%%%%%%%%%%%%%%%%%%%
\section{Constraining the degree of magnetization and corrugation of the solar TR}
\label{sec:S4}

We aim at constraining the magnetic field strength and degree of corrugation of the TR corresponding to the 
solar disk regions observed by CLASP. To this end, we confront the $Q/I$ and $U/I$ line-center signals measured 
by CLASP (hereafter, $\vec S^{\rm obs}$), with the theoretical line-center signals ($\vec S$) calculated in a grid of 3D models characterized by 
the scaling factors $\beta$ and $\gamma$, with $\beta=\gamma=1$ corresponding to the 3D model described in Carlsson et al. (2016). 
Each pixel along the 400 arcsec spectrograph slit image corresponds to a particular LOS, and we have considered those having $\mu$ values between 0.1 and 0.8, with the exception of the faint filament region located around $\mu=0.7$ in Fig. 1. The theoretical $I$, $Q$, and $U$ profiles have been degraded to mimic the CLASP resolution according to the laboratory measurements described in Giono et al. (2016). 

To determine the 3D model whose emergent $Q/I$ and $U/I$ line-center signals are the closest to those observed by CLASP, we apply the statistical inference approach discussed in Section 3.1 of \v{S}t\v{e}p\'an et al. (2018), which gives the following expression for the posterior of the hyper-parameters $\vec\theta$ (i.e., the corrugation parameter $\gamma$ and the magnetization parameter $\beta$) based on the line-center data at all the CLASP spatial pixels $i$: 
\begin{equation}
p(\vec\theta|\vec S^{\rm obs})=\prod_i \mathcal{L}_i(\vec\theta|\vec S^{\rm obs}) p(\vec\theta)=\mathcal{L}(\vec\theta|\vec S^{\rm obs})p(\vec\theta)\,,
\label{eq:finpost}
\end{equation}
where $p(\vec\theta)$ is the prior of the hyper-parameters. The likelihood $\mathcal{L}_i(\vec\theta|\vec S^{\rm obs})$ is

\begin{equation}
\mathcal{L}_i(\vec\theta|\vec S^{\rm obs})\equiv\int d\vec S\;\frac{p^{\mu_i}(\vec S|\vec\theta)}{(\sqrt{2\pi}\sigma)^2}
\exp\left\{ -\frac{\| \vec S-\vec S^{\rm obs} \|^2}{2\sigma^2} \right\}\,,
\label{eq:llii}
\end{equation}
with $\sigma^2$ the variance of the uncorrelated Gaussian noise. 
We point out that $p^{\mu_i}(\vec S|\vec\theta)$ are the hierarchical priors of $\vec S$ that depend on the hyper-parameters $\vec\theta$. 
%is the probability of $\vec S$ conditional to the truth of $\vec\theta$, 
This is derived from the histograms of the $\vec S$ signals calculated for each $\mu_i$ value in each particular 3D model characterized by its scaling factors $\beta$ and $\gamma$.
In the CLASP observations the standard deviation of the noise 
is $\sigma=0.05\%$ after averaging three pixels in wavelength around
the line center.
 
Our prior for the corrugation of the TR is that any compression of the 3D model is allowed, while   
expansions are much less likely (i.e., $p(\gamma)=1/2$ for $\gamma\in(0,1]$ and $p(\gamma)={\rm exp}(-(\gamma-1))/2$ for $\gamma>1$). 
For the magnetic field strength of the TR of the quiet Sun we have deemed reasonable 
to use $p(B)={\rm exp}(-B/130)/130$, based on the argument that the mean field strength of the quiet regions of the solar photosphere is of the order of 100 G and that 
it decreases with height in the atmosphere (Trujillo Bueno et al. 2004). This is our first step for determining the parametrization $\vec{\theta}$ of the above-mentioned 3D statistical model of the solar atmosphere that maximizes the marginal posterior of Eq. (5). Our aim is to estimate some global properties of the quiet Sun atmosphere observed by CLASP, i.e., the mean field strength and the degree of corrugation of the chromosphere-corona TR. 

Figure 4 shows the result.
It suggests that 3D models with less magnetization in the TR than in the model of Carlsson et al. (2016) produce scattering polarization signals in better agreement with those observed by CLASP. A more robust conclusion is that, among the magnetized models needed to explain the CLASP observations, 3D models with a significantly larger (i.e., $\gamma{\approx}1/3$) degree of corrugation of the TR would yield a better agreement with the observations. 

\begin{figure}[t]
\begin{center}
\includegraphics[scale=1.2]{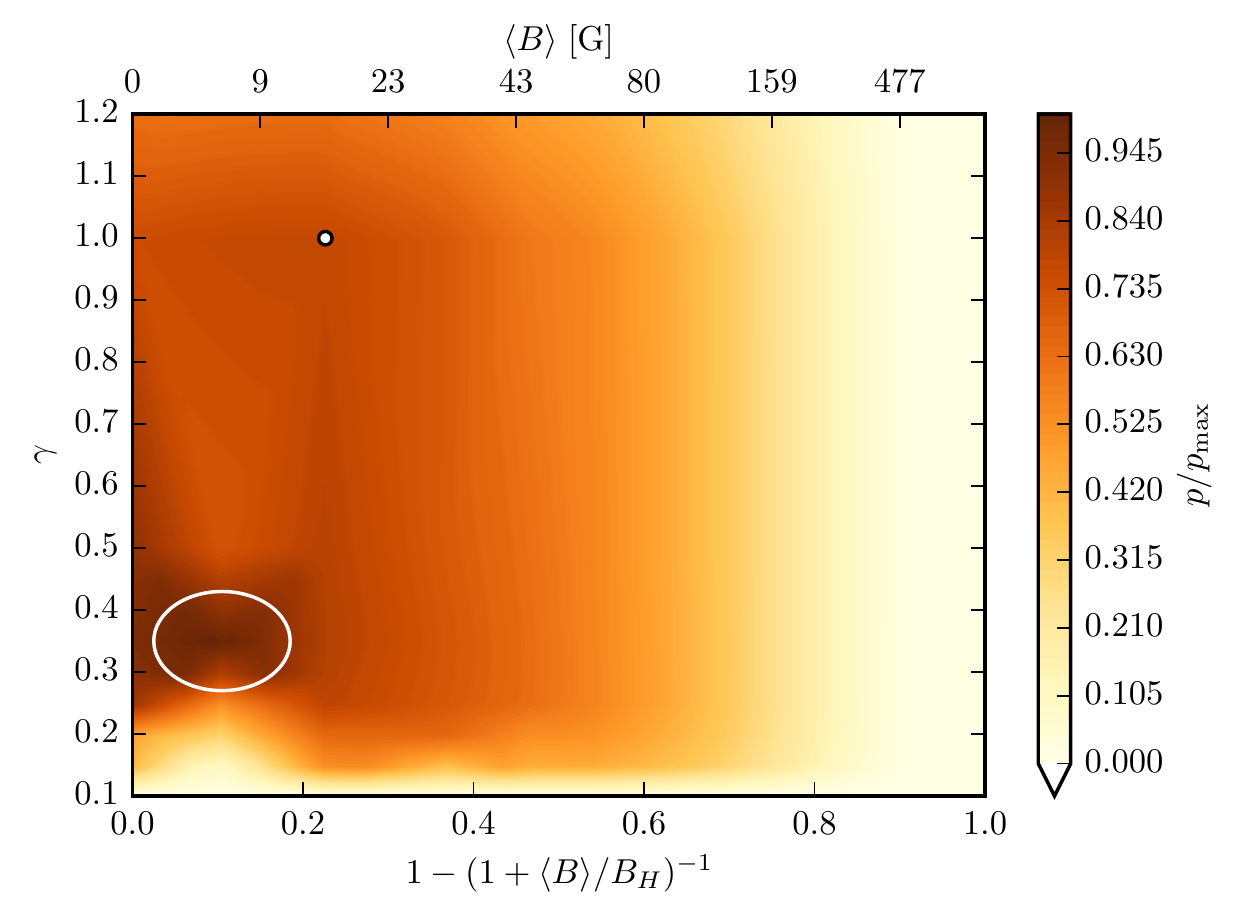}
\caption{The variation with $\beta$ and $\gamma$ of the marginal posterior of Eq. (5), 
which quantifies the ability of the parametrized 3D models discussed in the text to explain the Ly-$\alpha$ line-center data.
$B_H=53$ G is the line's Hanle critical field. Each value of $\beta$ implies a TR mean field strength ${\langle B \rangle}$.
The white circle corresponds to the 3D model of Carlsson et al. (2016). 
Note that the posterior peaks for ${\langle B \rangle}\,{\approx}\,7$ G and a compression factor $\gamma\approx {1/3}$.  
}
\label{figure-5}
\end{center}
\end{figure}

%%%%%%%%%%%%%%%%%%%%%%%%%%%%%%%%%%%%%%%%
%%%%%%%%%%%%%%%%%%%%%%%%%%%%%%%%%%%%%%%%
%%%%    SECTION 6
%%%%%%%%%%%%%%%%%%%%%%%%%%%%%%%%%%%%%%%%
%%%%%%%%%%%%%%%%%%%%%%%%%%%%%%%%%%%%%%%%
\section{Conclusions}
\label{sec:concl}

Our statistical approach to interpret the spectropolarimetric observations of the hydrogen Ly-$\alpha$ line achieved by CLASP suggests that the mean field strength of the TR of the observed quiet Sun regions is significantly lower than the 15 G of Carlsson et al's (2016) 3D model, perhaps not surprisingly since it is for an enhanced network region, and that a TR plasma with a substantially larger degree of geometrical complexity is needed to explain the CLASP observations.

We think that our conclusion on the geometric complexity of the TR plasma is more robust than the one on its degree of magnetization because in order to cancel completely the CLV of the $Q/I$ line-center signals we need to increase the degree of corrugation of the model's TR, while the information on its magnetic strength is mainly encoded in the peak-to-peak amplitudes of the $Q/I$ and $U/I$ spatial radial variations, which are sensitive to the CLASP point spread function used (based on laboratory measurements) and to the 3D model atmosphere chosen for the statistical inference. In any case, a key point to emphasize is that the linear polarization produced by scattering processes in the core of the hydrogen Ly-$\alpha$ line encodes valuable information on the magnetic field and geometrical complexity of the chromosphere-corona TR, which can be revealed with the help of statistical inference methods such as the approach applied in this paper. 
Clearly, having simultaneous spectropolarimetric observations in two or more spectral lines would facilitate the determination of the plasma magnetization, especially if the main difference between the spectral lines used lies within their sensitivity regime to the Hanle effect.

Evidently, we need more realistic 3D numerical models of the upper chromosphere of the quiet Sun. 
Chromospheric spicules are ubiquitous in subarcsecond resolution Ly-$\alpha$ filtergrams (Vourlidas et al. 2010), 
but they are not present in Carlsson et al's (2016) model. Our investigation suggests that 3D models with such needle-like plasma structures 
all over the field of view would be a much better representation of the geometric complexity of the TR plasma observed by CLASP. Finally, it is also important to emphasize that a suitable way to validate or refute numerical models of the chromosphere-corona TR is by confronting calculations and observations of the scattering polarization in ultraviolet lines sensitive to the Hanle effect. We plan to pursue further this line of research by also exploiting the polarization observed by CLASP in the resonance line of Si {\sc iii} at 1206 \AA\ (see Ishikawa et al. 2017) and, of course, the observations of the Mg {\sc ii} $h$ \& $k$ lines that the future flight of CLASP-2 will provide.

%%%%%%%%%%%%%%%%%%%%%%%%%%%%%%%%%%%%%%%%
%%%%%%%%%%%%%%%%%%%%%%%%%%%%%%%%%%%%%%%%
%%%%    ACKNOWLEDGEMENTS
%%%%%%%%%%%%%%%%%%%%%%%%%%%%%%%%%%%%%%%%
%%%%%%%%%%%%%%%%%%%%%%%%%%%%%%%%%%%%%%%%

\acknowledgements

The CLASP team is an international partnership between NASA Marshall Space Flight Center, 
National Astronomical Observatory of Japan (NAOJ), Japan Aerospace Exploration Agency 
(JAXA), Instituto de Astrof\'{i}sica de Canarias (IAC) and Institut d'Astrophysique Spatiale; 
additional partners are the Astronomical Institute ASCR, Istituto Ricerche Solari Locarno (IRSOL), 
Lockheed Martin and University of Oslo. The US participation was funded by NASA Low Cost Access to Space (Award 
Number 12-SHP 12/2-0283). The Japanese participation was funded by the basic research 
program of ISAS/JAXA, internal research funding of NAOJ, and JSPS KAKENHI Grant 
Numbers 23340052, 24740134, 24340040, and 25220703. The Spanish participation was 
funded by the Ministry of Economy and Competitiveness through project AYA2010-18029 
(Solar Magnetism and Astrophysical Spectropolarimetry). The French hardware participation 
was funded by Centre National d'Etudes Spatiales (CNES). Moreover, we acknowledge the grants provided by the Barcelona Supercomputing Center (National Supercomputing Center, Barcelona, Spain), as well as the funding received through grant \mbox{16--16861S} of the Grant Agency of the Czech Republic, project \mbox{RVO:67985815} of the Czech Academy of Sciences, and projects \mbox{AYA2014-60476-P} and \mbox{AYA2014-55078-P} of the Spanish Ministry of Economy and Competitiveness. JS is grateful to the Severo Ochoa Visiting Researchers Programme of the IAC for financing a few months working visit at the IAC. 
JTB acknowledges the funding received from the European Research Council (ERC) under the European Union's Horizon 2020 research and innovation programme (ERC Advanced Grant agreement No 742265).

\software{PORTA (\v{S}t\v{e}p\'an \& Trujillo Bueno 2013)}

%%%%%%%%%%%%%%%%%%%%%%%%%%%%%%%%%%%%%%%%
%%%%%%%%%%%%%%%%%%%%%%%%%%%%%%%%%%%%%%%%
%%%%    REFERENCES
%%%%%%%%%%%%%%%%%%%%%%%%%%%%%%%%%%%%%%%%
%%%%%%%%%%%%%%%%%%%%%%%%%%%%%%%%%%%%%%%%

\end{document}